\documentstyle[aps,preprint,tighten]{revtex}

\def\mmeins#1#2#3#4#5#6#7#8#9{\mathord{1\hskip #1
    \vrule width #2 height #3 depth #4 \hskip #5
    \vrule width #6 height #7 depth #8 \hskip #9}}
\def\meins{\mathchoice
 {\mmeins{-1.50pt} {0.50pt}{7.75pt}{-0.2pt}
         {-1.20pt} {2.50pt}{0.30pt}{-0.05pt} {1.50pt}}
 {\mmeins{-1.50pt} {0.50pt}{7.75pt}{-0.2pt}
         {-1.20pt} {2.50pt}{0.30pt}{-0.05pt} {1.50pt}}
 {\mmeins{-1.17pt} {0.40pt}{5.13pt}{-0.2pt}
         {-0.67pt} {1.67pt}{0.80pt}{-0.2pt} {1.00pt}}
 {\mmeins{-1.00pt} {0.30pt}{3.85pt}{-0.2pt}
         {-0.50pt} {1.35pt}{0.80pt}{-0.2pt} {0.75pt}}}

\def\G{{\cal G}}
\def\R{{\cal R}}
\def\H{{\cal H}}
\def\L{{\cal L}}
\def\ba{\begin{eqnarray}}
\def\ea{\end{eqnarray}}
\def\nn{\nonumber}
\def\tr{\mbox{tr}}
\def\e{\mbox{e}}
\def\d{\mbox{d}}

\begin{document}
\preprint{HD-TVP-94-17}

\title{{Rigorous mean field model for CPA: \\
         Anderson model with free random variables}}

\author{Peter Neu$^1$ and Roland Speicher$^{2*}$}
\address{$^1$ Institut f\"ur Theoretische Physik, Universit\"at Heidelberg,
Philosophenweg 19, 69120 Heidelberg, Germany}
\address{$^2$ Institut f\"ur Angewandte Mathematik, Universit\"at Heidelberg,
Im Neuenheimer Feld 294, 69120 Heidelberg, Germany}
\date{September 20, 1994}
\maketitle

\begin{abstract}
\noindent A model of a randomly disordered system with site-diagonal random
energy fluctuations is introduced.  It is an extension of Wegner's $n$-orbital
model to arbitrary eigenvalue distribution in the electronic level
space. The new feature is that the random
energy values are not assumed to be independent at different sites but free.
Freeness of random variables is an analogue  of  the concept
 of independence
for non-commuting random operators. A possible realization is the ensemble
of  at different lattice-sites randomly  rotated matrices.
The one- and  two-particle Green functions of the
 proposed  hamiltonian are calculated exactly.
 The  eigenstates are extended and the
 conductivity  is nonvanishing  everywhere inside the band.
The long-range behaviour and the zero-frequency limit of the
two-particle Green function are universal with respect to the
eigenvalue distribution in the electronic level space.
  The solutions  solve
the CPA-equation for the one- and two-particle Green function
of the corresponding   Anderson model. Thus our (multi-site)
model is a rigorous mean field model for the (single-site) CPA.
  We show how the Llyod model is included in our model and treat various kinds
of noises.

\end{abstract}
$\\$

{\bf Key-words}: Disordered systems, random matrices, CPA.
\pacs{PACS numbers: 71.10.+x, 05.60.+w, 71.55.-i}

\narrowtext

\section{Introduction} \label{S1}
During the last  decades  randomly disordered systems
have gained much interest  in statistical physics. Especially, since
Anderson's paper \cite{And} in 1958 these systems have attracted
many physisists due to the localization phenomenon. However, realistic
multi-site models, like the Anderson model, are in general unsolvable beyond
the
one-dimensional case.
Exact calculations are only possible in one pathological case -- namely for
the Lloyd model \cite{Llo} with Cauchy-distributed disorder.

The Anderson model describes the hopping of an electron in a $d$-dimensional
disordered  lattice ${\bf Z}^d$. The hamiltonian is
\ba  \label{I1}
H = H_0 + H_1\ ,
\ea
where $H_0$ is the deterministic, translational invariant hamiltonian
\ba  \label{I2}
H_0 = \sum_{r,r'\in Z^d} v_{|r-r'|} |r\rangle\langle r'|\ ,
\ea
and where the random disorder is assumed to be diagonal in the sites and
independent
between different sites, i.e.
\ba  \label{I3}
H_1 = \sum_{r\in Z^d} f_r |r\rangle\langle r|
\ea
with $f_r$ being identically distributed, independent random variables.
Although $H_1$ is a quite simple operator its relation to $H_0$ is complicated
and in
the present form the Anderson model is not exactly solvable.

To circumvent this dilemma two strategies have been developed so far:
One is to approximate the multi-site model by  single-site models
which can be solved exactly \cite{Econ,Lif,Sov,Tay,Vel}, the other is to
develop
models which become exactly solvable in the mean field limit of
infinite dimension $d$, infinite interaction range $R$ or infinite number
of angular momentum states $n$ at each lattice site \cite{Weg,KP}.
Among the former the most realistic ones are
those single-site models which apply the
{\it coherent-potential approximation} (CPA) \cite{Sov,Tay,Vel};
 among the latter  Wegner's {\it n-orbital model}
\cite{Weg} is most frequently studied.  Wegner's generalization
of the Anderson model consists of putting $n$ electronic states at each site
and describing the disorder by gaussian random matrices
in the electronic states.
 Whereas for $n=1$ this reduces to the  Anderson model
 with gaussian disorder, the opposite limit
$n\to\infty$ becomes exactly solvable. Interestingly, this solution
coincides with a CPA-solution of the Anderson model where the single-site
disorder is distributed according to Wigner's semi-circle law \cite{Wig,Arn}.

This very fact has gained much interest in the debate about
 the range of validity
of CPA and its connection with the mean field models. As mentioned
by Khorunzhy and Pastur \cite{KP}  the infinite $d$, $R$ and $n$ limits do not
coincide with CPA in general; however, they have  similiar properties.

 The starting point of our investigation is the following observation:
  The main reason for the difficulty in solving the Anderson model
is that the assumption of independence of the $f_r$ at different
sites  cannot be translated into a tractable relation between $H_0$ and $H_1$.
 Hence, in our approximation
of the Anderson model, we replace the assumption of independence
by a ``non-commutative" independence: we assume the $f_r$ to be {\it free}.
Freeness has been introduced in mathematics in the context
 of von Neumann algebras by Voiculescu \cite{Voi1}
and has been extended
to  non-commutative probability theory
by Voiculescu \cite{Voi1,Voi2,Voi3,VDN,Voi4} and  Speicher
\cite{Spe3,KSp,Spe2}.
The assumption of freeness will allow us to calculate
all physical quantities in our model exactly.

Freeness and random matrices
are intimately  connected with each other:
Arbitrary hermitian $n\times n$-matrices  randomly rotated against each other
 -- via unitary random matrices --  are in the limit $n\to\infty$ a
possible representation of free random variables.
\cite{Voi3,Spe1}.
{}From this point of view our model has a mean field character and
can be considered as a generalization
of Wegner's {\it n-orbital model} to arbitrary eigenvalue distributions in the
electronic level space.  In contrast to this,
Wegner was restricted  to the distribution of symmetric random matrices, i.e.
 in the limit $n\to\infty$ to Wigner's  semi-circular law \cite{Wig,Arn}.
 It is this last restriction which we will
show  to be the reason that CPA and the
 hitherto considered mean field models  coincide only
 for the semi-circle distribution.
 On the contrary, we can show that the solution of our model and
the solution   of the corresponding
Anderson model in CPA-approximation
 coincide always if the disorder is
distributed according to the same distribution in both cases.
 Thus our (multi-site)
model is a rigorous mean field model for the (single-site) CPA.

The long-range behaviour and the zero-frequency limit of the
two-particle Green function are universal with respect to the
eigenvalue distribution in the electronic level space. Independently
of the distribution of the
disorder we find  Wegner's result for the gaussian ensemble \cite{Weg}
that (i) eigenstates separated by an energy
$\omega$  are correlated over a length $L$ which diverges like
$|\omega|^{-1/2}$ for $\omega\to 0$, and that
(ii) the two-particle Green function  for energies in opposite halves of the
complex plane differing by $\omega$ approaches a constant for $d>2$,
diverges logarithmically for $d=2$ and like $|\omega|^{d/2-1}$ for $0\le d<2$.

The paper is organized as follows: In sect. \ref{S2}
 we introduce the concept of
free random variables and outline their connection with random matrices and
their description by non-crossing cumulants.
In sect. \ref{S3}
 we introduce our model and calculate the one- and two particle
Green function   and the conductivity exactly. Sect. \ref{S4}
 is devoted to the
connection of our model with CPA.
In sect. \ref{S5}  we discuss our
model for  various  kinds of disorder, and show how the Llyod model is
included in our model, and finally, in sect. \ref{S6},
 we summarize our main  results.

\setcounter{equation}{0}
\section{Freeness, random matrices, and non-crossing cumulants}\label{S2}
\subsection{The concept of freeness and random matrices}\label{S21}
The concept  of {\it freeness} was introduced by Voiculescu \cite{Voi1}
in order to treat non-commutative random variables in an analogous way
as commutative (classical) random variables are treated by
the concept of {\it independence}.
{}From an operational point of view independence and freeness are nothing but
rules for the calculation of mixed moments of random variables $X_1, X_2, ...$,
if the moments of all $X_r$ are given, separately. Thus, independence of the
$X_r$ means
\ba  \label{CF1}
\langle X_{r(1)} X_{r(2)} X_{r(3)} ... \rangle =
\langle \prod_{i: r(i) = 1}X_{r(i)} \rangle\ \langle \prod_{i: r(i) = 2}
X_{r(i)} \rangle\ ...\ .
\ea
Freeness replaces this now by the following rule:$\\$

{\sc Definition}: $X_1, X_2,...$ are {\it free}, if we have
for all $m\in$ {\bf N} and for all
polynomials $p_1(X), p_2(X), ...,p_m(X)$ of one variable $X$  that
\ba  \label{CF3}
\langle p_1(X_{r(1)}) p_2(X_{r(2)}) \ldots p_m(X_{r(m)})\rangle = 0
\ea
whenever
\ba  \label{CF2}
\langle p_k(X_{r(k))} \rangle = 0
\ea
for all $k = 1,\ldots, m$   and $r(k) \neq r(k+1)$
for all $k = 1,\ldots, m-1$ (i.e. consecutive indices are different).$\\$

First, one should convince oneself that this is  really a rule for calculating
all mixed moments of the $X_r$'s. Let us consider the case of two variables
$X=X_1$ and $Y=X_2$. For $\langle X Y \rangle $ the definition  yields that
$\langle X Y \rangle = 0$ if $\langle X  \rangle = \langle Y  \rangle  = 0$.
If $X$ and $Y$ have non-vanishing mean then by using (\ref{CF3})
for the polynomials $p_1(x) = X-\langle X  \rangle \meins$ and
$p_2(Y) = Y-\langle Y  \rangle \meins$
one easily finds $\langle X Y \rangle  =
\langle X  \rangle\  \langle Y  \rangle $.
Whereas this is the same result as for independent $X$ and $Y$ the calculation
of $\langle X Y X Y\rangle$ via $0=\langle (X-\langle X  \rangle \meins)
 (Y-\langle Y  \rangle \meins)$$(X-\langle X  \rangle \meins)
  (Y-\langle Y  \rangle \meins) \rangle$ yields
\ba  \label{CF4}
\langle X Y X Y\rangle = \langle X^2 \rangle \langle Y\rangle^2 +
                                      \langle X \rangle^2 \langle Y^2\rangle -
		\langle X \rangle^2 \langle Y\rangle^2
\ea
and shows thereby that independence ($\langle X Y X Y\rangle =
 \langle X^2 \rangle \langle Y^2\rangle $) and freeness are quite different
 concepts. Furthermore, freeness is really a non-commutative concept:
 If $X$, $Y$ are free, we have
 $\langle X X Y Y \rangle = \langle X^2 \rangle \langle Y^2\rangle $ which
shows,
 cf. (\ref{CF4}),
 that $X$ and $Y$ do not commute. Hence $X$ and $Y$ cannot be represented by
 classical $c$-number random variables.

 There exist a  canonical
 representation of free random variables by special kinds of random matrices:
 Let $U(n)$ be the ensemble  of unitary $n\times n$-matrices equipped with the
canonical invariant Haar measure. Take two deterministic $n\times n$-matrices
$A$ and $B$ (e.g. diagonal matrices) and rotate them against each other
randomly, i.e.
 $X := A$ and  $Y := u B u^{\dag}$ with $u\in U(n)$.
 Then, in the limit $n\to \infty$,
$X$ and $Y$ are free with respect to  $\langle n^{-1}
\tr [\ldots]\rangle_{\rm av}$, where $\langle \ldots\rangle_{\rm av}$
denotes the average over the ensemble of unitary matrices. Note that
$n^{-1}\tr [\ldots]$ gives the eigenvalue distribution
of our $n\times n$-matrices. This connection
between freeness and unitary random matrices was first discovered by
Voiculescu \cite{Voi3}, and further developed by Speicher \cite{Spe1}.
Another representation for free random variables with a special kind of
distributions by
deformed creation and annihilation operators
will be discussed in Sect. \ref{S5}.

Let us now check that the assumption of
 freeness of the $f_r$ in the hamiltonian
$H_1$  results in a definite relation between $H_0$ and $H_1$, namely they are
also free.$\\$

{\sc Theorem} 1: Let the hamiltonian $H$ be given  by (\ref{I1})-(\ref{I3}).
If the $f_1, f_2, ...$ are free with respect
to $\langle\ldots\rangle_{\rm ens}$,
then $H_0$ and $H_1$ are also free with respect to $\langle\ldots\rangle$.
Here $\langle\ldots\rangle  = \langle\langle r_0|\ldots | r_0\rangle
\rangle_{\rm ens}$, independent of $r_0$, and $\langle\ldots\rangle_{\rm ens}$
denotes the average over the disorder.

{\sc Proof}: Consider polynomials $p_1,p_2,\ldots$ and
$q_1,q_2,\ldots$ with $\langle p_i(H_0)\rangle = 0 =  \langle q_j(H_1)\rangle$
for all $i$, $j$.  Then we have to show that
\ba
\langle p_1(H_0) q_1(H_1) p_2(H_0) q_2(H_1) \ldots\rangle = 0
\ea
and
\ba
\langle  q_1(H_1) p_1(H_0)  q_2(H_1) p_2(H_0) \ldots\rangle = 0\ .
\ea
We only treat the first case, the second is analogous. Note that
$\langle r| q_j(H_1)| r' \rangle = \delta_{r,r'} q_j(f_r)$. Then
\ba
\langle p_1(H_0) q_1(H_1) p_2(H_0) q_2(H_1) \ldots\rangle  =
\langle\langle r_0| p_1(H_0) q_1(H_1) p_2(H_0) q_2(H_1) \ldots|r_0\rangle
\rangle_{\rm ens} \nn\\
=  \sum_{r(1),r(2),...} \langle\langle r_0| p_1(H_0)|r(1)\rangle
q_1(f_{r(1)}) \langle r(1) | p_2(H_0)|r(2)\rangle   q_2(f_{r(2)})
\ldots|r_0\rangle
\rangle_{\rm ens}\nn\\
=  \sum_{r(1),r(2),...} \langle r_0| p_1(H_0)|r(1)\rangle
 \langle r(1) | p_2(H_0)|r(2)\rangle \ldots
 \langle  q_1(f_{r(1)}) q_2(f_{r(2)}) \ldots\rangle_{\rm ens}\ .
\ea
Since with $H_0$ also $p_i(H_0)$ is translationally invariant,
$\langle r_0| p_i(H_0)|r_0\rangle \equiv \langle p_i(H_0) \rangle = 0$
implies $\langle r(i)| p_{i+1}(H_0)|r(i)\rangle  = 0$ for all $i$, and we can
restrict
the sum to $r(i)$'s with $r(i) \neq r(i+1)$ for all $i$.
However for these terms
we know that $\langle  q_1(f_{r(1)}) q_2(f_{r(2)}) \ldots\rangle_{\rm ens} = 0$
due to the freeness of the $f_j$'s and $\langle  q_j(f_{r(j)}) \rangle_{\rm
ens}
= \langle  q_j(H_1) \rangle = 0$.  \hfill $\Box$

\subsection{Description of free random variables by non-crossing cumulants}
In the physics of disordered systems, usually, Green functions
 are  calculated. This leads to the
evaluation  of  mixed moments of  -- in our case -- free random variables.
 The abstract definition of freeness ensures that all mixed moments are
determined  but we do not have a concrete formula for them, so far.
An efficient machinery for concrete calculations are the non-crossing
cumulants.

Let $X_1, X_2, \ldots$ be free random variables.  Then, we consider  quantities
$k_m(Y_1,\ldots,Y_m)$ for all $m\ge 1$, where the arguments $Y_i$
are non-commutative polynomials in $X_1,X_2,\ldots$.  These $k_m$ are
called {\it non-crossing cumulants} and one way to define them is the following
recurrence formula between the moments and the cumulants
\ba  \label{NC1}
\langle Y_1\ldots Y_m\rangle &=& \sum_{p=0}^{m-1}
\sum_{i(1),\ldots,i(p) \atop \subset \{ 2,\ldots,m\} } k_{p+1}(Y_1,Y_{i(1)},
Y_{i(2)},\ldots, Y_{i(p)})\nn\\
&& \times  \langle Y_2\ldots Y_{i(1)-1}\rangle\langle Y_{i(1)+1}\ldots
 Y_{i(2)-1}\rangle \ldots \langle Y_{i(p)+1}\ldots Y_m\rangle\ .
\ea
Starting with $k_1(Y_1) = \langle Y_1\rangle$, (\ref{NC1}) may be used to
determine
$k_m(Y_1,\ldots,Y_m)$, succesively. The non-crossing cumulants were
introduced in \cite{Spe2} and further elaborated with regard to stochastic
dynamics in \cite{NSp1}. Examples  are (in an obvious notation)
$k_2(1,2) = \langle 1 2 \rangle - \langle 1 \rangle \langle 2 \rangle$;
 for the special case of centered $Y_i$ ($\langle Y_i \rangle = 0$
for $i=1,\ldots,4$), we have  $k_4(1,2,3,4) =$
$\langle  1 2 3 4  \rangle$ - $\langle 1 2 \rangle
\langle 3 4 \rangle$ - $\langle 1 4 \rangle \langle 2 3 \rangle$.

It follows from the results of \cite{Spe2} that  also  the following
generalization of (\ref{NC1}) holds
\ba  \label{NC2}
\langle Y_1 &\ldots & Y_m Y_{m+1}\ldots Y_{m+\ell}\rangle \  =\
\langle Y_1\ldots Y_m \rangle \langle Y_{m+1}\ldots Y_{m+\ell}\rangle\nn\\
&+& \sum_{p=1}^m \sum_{q=1}^\ell \sum_{i(1),\ldots,i(p)
\atop \subset \{ 1,\ldots,m\} }
\sum_{j(1),\ldots,j(q)
\atop \subset \{ m+1,\ldots,m+\ell\} }
k_{p+q}(Y_{i(1)},\ldots, Y_{i(p)},Y_{j(1)},\ldots, Y_{j(q)})\nn\\
&\times &  \langle Y_1\ldots Y_{i(1)-1}\rangle\langle Y_{i(1)+1}\ldots
 Y_{i(2)-1}\rangle \ldots \langle Y_{j(q)+1}\ldots Y_{m+\ell}\rangle\ .
 \ea
 Thus, the non-crossing cumulants give the corrections to the
 frequently assumed factorization
 of $\langle Y_1\ldots Y_m Y_{m+1}\ldots Y_{m+\ell}\rangle$
into $\langle Y_1\ldots Y_m \rangle \langle Y_{m+1}\ldots Y_{m+\ell}\rangle$.

To derive the connection between  freeness and non-crossing cumulants
we will use another characterization of the non-crossing cumulants
(which is equivalent to (\ref{NC1},\ref{NC2})), namely, they are uniquely
determined by
\ba   \label{NC3}
 k_1(Y) &=& \langle Y \rangle\\
         \label{NC4}
  k_m(Y_1,\ldots, Y_i , Y_{i+1},\ldots, Y_{m}) &=&
  k_{m-1}(Y_1,\ldots, Y_i Y_{i+1},\ldots, Y_{m}) \nn\\
  - \sum_{k=0}^{i-1} k_{m+k-i}(Y_1,\ldots, Y_k, Y_{i+1},&\ldots&, Y_{m})
  k_{i-k}(Y_{k+1},\ldots, Y_{i})\nn\\
  - \sum_{\ell=i+1}^{m-1}  k_{m+i-\ell}(Y_1,\ldots, Y_i, Y_{\ell+1},
&\ldots&, Y_{m})
  k_{\ell-i}(Y_{i+1},\ldots, Y_{\ell}).
\ea
Note that  in the first term on the rhs of (\ref{NC4}) $Y_i$ and  $Y_{i+1}$ are
multiplied
with each other, and that in the second and third term $Y_{k+1},\ldots, Y_{i}$
and $Y_{i+1},\ldots, Y_{\ell}$ are skipped in $k_{m+k-i}$ and $k_{m+i-\ell}$,
respectively.
Eq. (\ref{NC4}) allows to reduce all higher cumulants to $k_1$. It is quite
easy
to derive from (\ref{NC3},\ref{NC4}) the following properties:$\\$

\noindent {\sc Remarks}:\\
1. $k_m(Y_1,\ldots,Y_m)$  is a multi-linear function in $Y_1,\ldots,Y_m$.\\
2. $k_m(Y_1,\ldots,Y_m) = 0$ for $m\ge 2$ if at least for one $i$ we have
           $Y_i = \meins$ (of course $k_1(\meins) = 1)$.$\\$

Up to now we have not used any freeness; we have just defined non-crossing
cumulants as special polynomials in the moments. That this definition  gives us
indeed the right tool for handling freeness shows the next proposition
(for a more detailed proof see \cite{Spe2}):$\\$

{\sc Proposition 1}: For each $i=1,\ldots,m$
let $Y_i$ be a polynomial in one variable $X_{r(i)}$
for some $r(i)$, $Y_i = p_i(X_{r(i)})$,
and assume $X_1,X_2,\ldots$ to be free.
Then, $k_m(p_1(X_{r(1)}),\ldots,p_m(X_{r(m)})) = 0$
whenever  there exists at least one pair $i,j$ with $i\neq j$ and $r(i)\neq
r(j)$
 (i.e. such that $Y_i$ and $Y_j$ are free).

{\sc Proof}: By (\ref{NC4}) we can glue together neighbouring
$Y_i,Y_{i+1}$ with $r(i) =  r(i+1)$ and hence we can assume that $r(i)\neq
r(i+1)$
for all $i=1,\ldots, m-1$. Next, we write again
$Y_i = (Y_i - \langle Y_i \rangle \meins) +
\langle Y_i \rangle \meins$ and, by Remark 1 and 2, we can restrict
ourselves to the case where all $Y_i$ are centered,
 i.e. $\langle Y_i \rangle = 0$
for all $i=1,\ldots, m$. But then we can reduce --
by using (\ref{NC4}) (or equivalently (\ref{NC1})) and induction --
$k_m(Y_1,\ldots,Y_m)$ to $k_1(Y_1\dots Y_m)\equiv \langle Y_1\dots Y_m
\rangle$, which vanishes by the definition of the freeness. \hfill $\Box$ $\\$

Thus, the quite implicit definition of freeness, namely that very special
mixed moments in free variables vanish, has now been replaced by the statement
that all non-crossing cumulants with at least two different
free variables  vanish without any restriction on
$\langle p_i(X_{r(i)}) \rangle$ (cf. the definition of freeness).
So we have e.g. $k_3(X_1,X_1^2,X_2) = 0$ if $X_1$ and $X_2$ are free
-- independent of the values of $\langle X_1\rangle$, $\langle X_1^2\rangle$
and $\langle X_2\rangle$ -- whereas $k_3(X_1,X_1^2,X_1) \neq 0$ in general.
 It is this very property
of free random variables which will allow us to calculate the
one-particle and the two-particle Green function of our model in the next
section exactly.
  Note that the non-crossing cumulants play exactly the same role for free
  random variables as the usual cumulants do for independent random variables.

\setcounter{equation}{0}
\section{Site-diagonal Anderson model and freeness}  \label{S3}
\subsection{The model}  \label{S31}
We consider now the following model of a randomly disordered system:
The hamiltonian $H$ is given by $H = H_0 + H_1$, where $H_0$ and $H_1$
are defined in (\ref{I2}) and (\ref{I3}), respectively, and where we assume the
$f_r$ to be identically distributed and
free with respect to the average $\langle\ldots\rangle_{\rm ens}$.
   Due to  our discussion on the relation between freeness and
random matrices we also have the following concrete realization of the model:
At each lattice site $r$ we have $n$ electronic levels $|r\alpha\rangle$
numbered by $\alpha = 1,\ldots,n$. Let
$f = (\langle\alpha|f|\beta\rangle)_{\alpha,\beta=1}^n$ be a fixed operator in
the
electronic level space and put
\ba  \label{M1}
f_r := u_r f u_r^{\dag}
\ea
where the $u_r$ are unitary random matrices in the electronic level space
chosen independently for different sites $r$. This means that we act at each
site $r$
with a copy $f_r$ of the given operator $f$, but that the basis for $f_r$ and
the basis for
$f_{r'}$ are rotated randomly against each other for all pairs of different
sites
$r\neq r'$. According to our remarks around eq. (\ref{CF4}),
 the $f_r$ become free  in the limit $n\to\infty$.
Thus, freeness is the correct mathematical notion for the $n\to\infty$-limit.

Obviously, our model is nothing but a generalization of Wegner's
{\it n-orbital model}
\cite{Weg}.  In his original formulation Wegner has chosen the
$f_r = ((1/\sqrt{n})f_r^{\alpha\beta})_{\alpha,\beta=1}^n$
as gaussian random
matrices at each $r$, such that the entries $f_r^{\alpha\beta}$ and
$f_{r'}^{\gamma\delta}$ are independent for different sites $r\neq r'$.
Since random rotations between different sites do not alter this ensemble,
it follows that also in this model the $f_{r}$'s become free in the limit
$n\to\infty$. Hence Wegner's model is included as a special case in our model.

One should however realize that our model is much more general than Wegner's,
since
the gaussian ensemble corresponds to choosing a matrix with Wigner's
semi-circular eigenvalue distribution \cite{Wig,Arn}  for $f$ in (\ref{M1}).
In contrast to that we are totally free in choosing any eigenvalue distribution
for  $f$  and hence for $H_1$.

It is interesting to note that freeness is already to some extend contained in
the original Anderson model ($f_r$ being independent, random $c$-numbers).
For instance we find in this case for $\langle H_0 H_1 H_0 H_1\rangle$ the
 same result as  in (\ref{CF4}):
\ba  \label{M2}
\langle H_0 H_1 H_0 H_1\rangle &=&
\langle\langle r_0| H_0 H_1 H_0 H_1|r_0\rangle\rangle_{\rm ens}\nn\\
&=& \left\langle  \sum_r  v_{|r-r_0|} f_r  v_{|r_0-r|} f_{r_0}
\right\rangle_{\rm ens}\nn\\
&=& \sum_{r\neq r_0}  v_{|r-r_0|}  v_{|r_0-r|} \langle f_r\rangle_{\rm ens}
 \langle  f_{r_0}\rangle_{\rm ens} + v_0^2 \langle  f_{r_0}^2\rangle_{\rm
ens}\nn\\
 &=& \langle  H_1\rangle^2 \Big(\sum_{r}  v_{|r-r_0|}  v_{|r_0-r|}  -
v_0^2\Big)
 +  \langle  H_1^2\rangle v_0^2\nn\\
 &=& \langle  H_1\rangle^2  \langle  H_0^2\rangle +
 \langle  H_1^2\rangle \langle  H_0\rangle^2 -
 \langle  H_1\rangle^2  \langle  H_0\rangle^2 \ .
 \ea
 Thus, the usual Anderson model
 yields freeness between $H_0$ and $H_1$ for small moments, but something
 uncontrollable for higher moments which precludes
 the model to be exactly solvable.

 \subsection{One-particle Green function} \label{S32}
 We want to calculate the averaged one-particle Green function (1PG) defined by
 \ba  \label{1G1}
 G(r,r';z) := \left\langle\left\langle r\left| \frac{1}{z -
H}\right|r'\right\rangle
 \right\rangle_{\rm ens}\ .
 \ea
 In matrix notation this reads $G(r,r';z) = n^{-1}\sum_\alpha $$
 \langle\langle r,\alpha | [z-H]^{-1} | r',\alpha\rangle\rangle_{\rm ens}$.
 Let us first concentrate on its diagonal part $G(r_0,r_0;z)$ which is
independent
 of $r_0$ due to  translation invariance. Let us introduce the short-hand
 notation $G(z) := G(r_0,r_0;z)$  and $\langle\ldots \rangle :=
 \langle\langle r_0 | \ldots | r_0\rangle \rangle_{\rm ens}$, thus
 \ba  \label{1G2}
 G(z)  =  \left\langle  \frac{1}{z - (H_0 + H_1)}\right\rangle
 = \sum_{n=0}^\infty \frac{\langle (H_0 + H_1)^n\rangle }{z^{n+1}}\ .
 \ea
 On calculating this quantity  we assume that we know the 1PG of $H_0$ and
$H_1$,
 separately,
  \ba     \label{1G3}
 G_0(z)  &=&  \left\langle  \frac{1}{z - H_0}\right\rangle
 = \sum_{n=0}^\infty \frac{\langle H_0^n\rangle }{z^{n+1}}
  \equiv  \sum_{n=0}^\infty \frac{\langle r_0 | H_0^nJ| r_0 \rangle
}{z^{n+1}}\\
            \label{1G4}
 G_1(z)  &=&  \left\langle  \frac{1}{z -  H_1}\right\rangle
 = \sum_{n=0}^\infty \frac{\langle H_1^n\rangle }{z^{n+1}}
  \equiv \sum_{n=0}^\infty
\frac{\langle f_{r_0}^n\rangle_{\rm ens} }{z^{n+1}}\ .
 \ea
 Thus,  our problem  consists  of    calculating
  moments of $H_0 + H_1$ given the moments of $H_0$ and $H_1$, separately,
  where due to our assumption of the $f_r$ being free and by
  Theorem 1,  $H_0$ and $H_1$ are free.
 In analogy to the usual convolution, which describes the sum of independent
 random variables,  we have to calculate the so called {\it free convolution}
 \cite{Voi2,VDN} of the free random variables $H_0$ and $H_1$.
 At this point the difficulties with the usual Anderson model become evident:
 Independence of  $f_1, f_2,\ldots$ does not imply a definite relation between
 $H_0$ and $H_1$ so that no well-defined notion of convolution between the
 distributions of $H_0$ and $H_1$ exists. $\\$

 {\sc Theorem} 2: Let the hamiltonian $H$ be given
  by (\ref{I1})-(\ref{I3}) where
 $f_1,f_2,\ldots$ are free and identically  distributed.
Then, the diagonal part of  the 1PG is given by
 \ba  \label{1G12}
G(z) = G_0\Big[z - R_1[G(z)]\Big]
\ea
where $R_1$ is determined by
\ba         \label{1G9}
G_1(z) &=& \frac{1}{z - R_1[G_1(z)]}\ .
\ea
The off-diagonal part of the 1PG is given via  the Fourier transform
\ba  \label{1G16}
G(r,r';z)  = \int_q \tilde{G}(q;z)\, \e^{iq(r-r')}
\ea
by
\ba  \label{1G20}
\tilde{G}(q;z) =\frac{1}{z - \tilde{v}(q) - R_1[G(z)]}\ ,
\ea
where $v_{r-r'} = \int_q \tilde{v}(q)\, \e^{iq(r-r')}$. Here,
$\int_q := \frac{{\cal V}}{(2\pi)^d}\int_{1BZ} \d^dq$,
${\cal V}$ being the volume of the first Brillouin zone ($1BZ$).
$\\$

\noindent {\sc Remark}: \\
Note that the diagonal part $G(z)$ entirely determines the off-diagonal part
since with $\tilde G_0(q;z) = [z-\tilde v(q)]^{-1}$, eqs. (\ref{1G16}) and
{\ref{1G20})
are equivalent to
\ba  \label{1G20X}
G(r,r';z) = G_0(r,r';z - R_1[G(z)])\ .
\ea
$\\$

 {\sc Proof}: We first prove Theorem 2 for the  diagonal part of the 1PG.
 Proposition 1 implies that the non-crossing cumulants
of $H = H_0 + H_1$ are  additive, i.e.
 \ba  \label{1G5}
 k_m(H) = k_m(H_0) + k_m(H_1)
 \ea
 with the short-hand notation $ k_m(H) :=  k_m(H,\ldots,H)$. Hence the free
 convolution is linearized by the non-crossing cumulants as the usual
convolution
 is linearized by the usual cumulants.
It remains to derive a relation between the
  non-crossing cumulants and the 1PG.
  If we specialize (\ref{NC1}) to $Y_1 = \ldots = Y_m = H$ we obtain
\ba  \label{1G6}
\langle  H^m\rangle  =  \sum_{p=1}^m
\sum_{j(1),\ldots,j(p) = 0 \atop j(1)+\ldots+j(p) = m-p}^{m-p} k_{p}( H )
  \langle  H^{j(1)}\rangle  \ldots \langle  H^{j(p)}\rangle\ .
\ea
If we now define
\ba  \label{1G7}
R(w) := \sum_{m=0}^\infty w^m k_{m+1}(H)
\ea
then we find with (\ref{1G2},\ref{1G6}) the relation
\ba  \label{1G8}
G(z) = \frac{1}{z - R[G(z)]}\ .
\ea
 Thus, $R$ can be considered as the  self-energy of $G(z)$ which depends
 self-consistently on $G(z)$ itself. Relation (\ref{1G8}) and its equivalent
 form
 \ba  \label{VoiRel}
  G[R(w) + w^{-1}] = w
  \ea
  are  due to Voiculescu \cite{Voi2} who calls $R$ the
 $R$-transform of $H$. The derivation given here using non-crossing cumulants
 has first been given in \cite{Spe2}, for a dynamical generalization see
\cite{NSp1}.
 In the same way we can write
 \ba  \label{1G9a}
G_0(z) &=& \frac{1}{z - R_0[G_0(z)]} \\
        \label{1G9b}
G_1(z) &=& \frac{1}{z - R_1[G_1(z)]}
\ea
where
\ba  \label{1G10a}
R_0(w) &:=& \sum_{m=0}^\infty w^m k_{m+1}(H_0) \\
        \label{1G10}
R_1(w) &:=& \sum_{m=0}^\infty w^m k_{m+1}(H_1)\ .
\ea
Because of (\ref{1G5}) $R$ is also additive
\ba  \label{1G11}
R(w) = R_0(w) + R_1(w)\ .
\ea
Defining $y$ by $G(z) = G_0(y)$ we get
$z - R_0[G(z)] - R_1[G(z)]$ $= y - R_0[G_0(y)]$
$= y - R_0[G(z)]$ and thus
$z - R_1[G(z)] = y$ from which we finally derive (\ref{1G12})
which proves together with (\ref{1G9b})  the first assertion of our theorem.
 Note that (\ref{1G12}) reduces to (\ref{1G9}) if we put $H_0 = 0$: then
 $G_0(z) = z^{-1}$ and $G = G_1$. Note also that there is only an
 apparent asymmetry between
 $H_0$ and $H_1$, since we may write in the same way
 \ba  \label{1G13}
G(z) = G_1\Big[z - R_0[G(z)]\Big]\ .
\ea
 Let us now treat the off-diagonal part of the 1PG.
 Again, our assumption of freeness
 of the $f_r$ will guarantee that we can derive
an exact expression for $G(r,r',z)$.
Using Dyson's equation gives
\ba  \label{1G14}
G(r,r';z) &=& G_0(r,r';z) \nn\\
&+& \sum_{m=1}^\infty\sum_{r_1,\ldots,r_m}
\langle G_0(r,r_1;z) f_{r_1} G_0(r_1,r_2;z) f_{r_2} \ldots
 f_{r_m} G_0(r_m,r';z)\rangle_{\rm ens}\nn\\
 &=&  G_0(r,r';z) + \sum_{m=1}^\infty\sum_{r_1,\ldots,r_m}
 G_0(r,r_1;z)    \ldots   G_0(r_m,r';z)
 \langle   f_{r_1} \ldots  f_{r_m}  \rangle_{\rm ens}\nn\\
 &=& G_0(r,r';z) + \sum_{m=1}^\infty\sum_{r_1,\ldots,r_m}
 G_0(r,r_1;z)    \ldots   G_0(r_m,r';z) \nn\\
 && \times \sum_{p=0}^{m-1}
\sum_{i(1),\ldots,i(p) \atop \subset \{2,\ldots,m\}}
k_{p+1}(f_{r_1},f_{r_{i(1)}},
\ldots, f_{r_{i(p)}})\nn\\
&& \times  \langle f_{r_2} \ldots f_{r_{i(1)-1}}\rangle\langle
f_{r_{i(1)+1}}\ldots
 f_{r_{i(2)-1}}\rangle \ldots \langle f_{r_{i(p)+1}}\ldots f_{r_m}\rangle
 \ea
 where we have used the recurrence formula (\ref{NC1}) for the non-crossing
 cumulants. Due to the freeness of the $f_r$ and Proposition 1
 only such terms contribute where
   $r_1 = r_{i(1)} = \ldots = r_{i(p)}$, which yields after some resummations
 \ba  \label{1G15}
 G(r,r';z) &=& G_0(r,r';z) \nn\\
&+& \sum_{p=1}^\infty\sum_{r_1}  G_0(r,r_1;z) k_{p+1}(f_{r_1}) G(r_1,r_1;z)^p
G(r_1,r';z)\nn\\
&=& G_0(r,r';z)  + R_1[G(z)] \sum_{r_1}  G_0(r,r_1;z)  G(r_1,r';z)
\ea
where we have used $k_p(f_{r_1}) = k_p(H_1)$ and  $G(z) = G(r_1,r_1;z)$.
Note that, once $G(z)$ is known,
(\ref{1G15}) is a linear system of equations for $G(r,r';z)$ which can be
solved by
Fourier transformation.
 Then,  eq. (\ref{1G15})
reads
\ba  \label{1G18}
\tilde{G}(q;z) = \tilde{G}_0(q;z) + R_1[G(z)] \tilde{G}_0(q;z) \tilde{G}(q;z)
\ea
which yields with
\ba  \label{1G19}
\tilde{G}_0(q;z) = \frac{1}{z - \tilde{v}(q)}\ ,
\ea
 the second assertion eq. (\ref{1G20}) of our theorem.
 \hfill $\Box$
$\\$

We finally comment on the analytic structure  of the solution (\ref{1G12}).
By definition $G(z)$, $G_0(z)$ and $G_1(z)$ are holomorphic functions in the
upper complex half plane ${\bf C}^+$. However, it is a priori not clear whether
$G(z)$ initially defined by eq. (\ref{1G12}) only in a
neighbourhood of $\infty$  has an analytic continuation to ${\bf C}^+$.

Clearly, the implicit definition of $G(z)$ by (\ref{1G12}) is unique
except for the critical points $z\in {\bf C}^+$ where $G'(z) = 0$.
Let us denote this set by $D = \{ z\in {\bf C}^+|G'(z) = 0\}$ and
by $\Delta = G(D)$ their critical values; analogously, we define
$D_j$ and $\Delta_j$ for $j=0,1$. Then Voiculescu has shown in \cite{Voi4}
for compactly supported measures of $H_0$ and $H_1$ and $G(z)$
implicitly given by (\ref{1G12}) that
\begin{enumerate}
\item $G({\bf C}^+) \subset G_0({\bf C}^+) \cap G_1({\bf C}^+)$.
\item $R_j[G_j(z)]$, $(j = 0,1,2)$
      has an analytic continuation from $\infty$
      to ${\bf C}^+$ (here $G_2 \equiv G$,
      $R_2 \equiv R$).
\item if $\Delta_0 \cap \Delta_1 = \emptyset$, the function
      $y(z) :=  z - R_1[G(z)]$ has an analytic continuation from
      $\infty$ to ${\bf C}^+$.
\end{enumerate}
The assumption $\Delta_0 \cap \Delta_1 = \emptyset$ implies that if
$w^{-1} + R_0(w)$
-- which is the inverse of $G_0$ (cf. eq. (\ref{VoiRel})) --
has a branching point of order $p > 1$ at $\zeta_0$ then $\zeta_0$
is not a branching point of $w^{-1} + R_1(w)$ and hence $\zeta_0$ is a
branching
point of order exactly $p$ for $w^{-1} + R(w)$ \cite{Voi4};
 this guarantees the
uniqueness of the analytic continuation of eq. (\ref{1G12}).
This assumption is not restrictive for practical purposes in physics
as long as the measures of $H_0$ and $H_1$ do not coincide. However,
 in this case
we know due to the freeness of $H_0$ and $H_1$ (cf. Theorem 1) that
$R(w) = 2 R_0(w) = 2 R_1(w)$.

Before we develop further our formalism for the 2PG, let us give $R_1[G(z)]$
a clear physical interpretation. The diagrammatic representation of eq.
(\ref{1G15}) is
\begin{center} \fbox{here: Diagram 1}\end{center}
in obvious notation.  Thus,  $R_1[G(z)]$ behaves like an effective local
potential   which scatters the propagating
electron  incoherently at each lattice site.  The total scattered wave
is the sum of the contributions from each lattice site without interference
terms.
 From this viewpoint our model has CPA character. We will come back to this
 in sect. \ref{S4}.

\subsection{Two-particle Green function} \label{S33}
In this section the averaged two-particle Green function (2PG)
\ba  \label{2G1}
\G (r,s,s',r';z_1,z_2) := \left\langle\left\langle r\left| \frac{1}{z_1 -
H}\right|s\right\rangle
\left\langle s'\left| \frac{1}{z_2 - H}\right|r'\right\rangle
\right\rangle_{\rm ens}
\ea
will be calculated.
In matrix notation this reads $\G (r,s,s',r';z_1,z_2) =
n^{-1}\sum_{\alpha,\beta}$$
 \langle\langle r,\alpha | [z_1-H]^{-1} | s,\beta\rangle
 \langle s',\beta | [z_2-H]^{-1} | r',\alpha\rangle\rangle_{\rm ens}$.
We define the 2PG
of $H_1$  by
\ba  \label{2G7}
\G_1(z_1,z_2) := \left\langle  \frac{1}{z_1 - H_1}
\frac{1}{z_2 - H_1}  \right\rangle  = \left\langle  \frac{1}{z_1 - f_{r_0}}
\frac{1}{z_2 - f_{r_0}}  \right\rangle_{\rm ens}\ .
\ea
{}From the identity
\ba
(z_1 - z_2)& & \sum_s \left\langle\left\langle r\left| \frac{1}{z_1 - H}
\right|s\right\rangle
\left\langle s\left| \frac{1}{z_2 - H}\right|r'\right\rangle \right\rangle_{\rm
ens}\nn\\
&=& \left\langle\left\langle r\left| \frac{z_1 - z_2}{(z_1 - H)(z_2 - H)}
\right|r'\right\rangle \right\rangle_{\rm ens}\nn\\
&=& G(r,r';z_2) - G(r,r';z_1)
\ea
one obtains the {\it sum rule}
\ba  \label{2G10}
(z_1 - z_2) \sum_s \G (r,s,s,r';z_1,z_2) =  G(r,r';z_2) - G(r,r';z_1)\ ,
\ea
 which reduces for $\G_1$ to
\ba  \label{2G11}
(z_1 - z_2) \, \G_1 (z_1,z_2) =  G_1(z_2) - G_1(z_1)\ .
\ea
The 2PG of our model is now given by the following $\\$

{\sc Theorem} 3: Let the hamiltonian $H$ be given  by (\ref{I1})-(\ref{I3})
where
 $f_1,f_2,\ldots$ are free and identically  distributed. Let further $G(z)$ and
 $G(r,r';z)$ be given by (\ref{1G12})-(\ref{1G20}). Then, for Im $z_{1/2} \neq
0$,
 \ba \label{2G6}
\G (r,s,s',r';z_1,z_2) &=& G(r,s;z_1) G(s',r';z_2)\nn\\
&+&  \R_1[G(z_1), G(z_2)] \nn\\
&\times & \sum_{r''} G(r,r'';z_1)  \G (r'',s,s',r'';z_1,z_2)   G(r'',r';z_2) \
,
\ea
where $\R_1(w_1,w_2)$
 is determined by
 \ba \label{2G12}
\R_1(w_1,w_2) = \frac{R_1(w_1) - R_1(w_2)}{w_1 - w_2}\ .
\ea

\noindent{\sc Remark}:\\
 Note that, once the 1PG is known,
(\ref{2G6}) is  a linear system of equations for the 2PG.

 {\sc Proof}: Again we use Dyson's equation to obtain
\ba  \label{2G3}
\G (r,s,s',r';z_1,z_2)  =
\left\langle\left\{ G_0(r,s;z_1) + \sum_{m=1}^\infty \sum_{r_1,\ldots,r_m}
G_0(r,r_1;z_1) f_{r_1} \ldots f_{r_m} G_0(r_m,s;z_1) \right\}\right.\nn\\
\times\   \left.\left\{ G_0(s',r';z_2)
 +   \sum_{\ell=1}^\infty  \sum_{s_1,\ldots,s_\ell}
G_0(s',s_1;z_2) f_{s_1} \ldots f_{s_\ell} G_0(s_\ell,r';z_2)
\right\}\right\rangle_{\rm ens} \ .
\ea
As before we can express this in terms of the non-crossing cumulants by using
relation (\ref{NC2}) for $\langle f_{r_1} \ldots f_{r_m} f_{s_1} \ldots
f_{s_\ell}\rangle$.
Due to the freeness of $f_1,f_2,\ldots$ we can  again use Proposition 1 and
therefore
we can restrict the summation in (\ref{NC2}) to terms with
$r_{i(1)} = \ldots = r_{i(p)}$ $ = s_{j(1)} = \ldots = s_{j(q)}$.
After some resummation we finally obtain
\ba  \label{2G4}
\G (r,s,s',r';z_1,z_2) &=&  G(r,s;z_1) G(s',r';z_2)\nn\\
&+& \sum_{p,q = 1}^{\infty} \sum_{r''} k_{p+q}(H_1)  G(r,r'';z_1)
G(r'',r'';z_1)^{p-1}\nn\\
& \times & \G (r'',s,s',r'';z_1,z_2) G(r'',r'';z_2)^{q-1} G(r'',r';z_2) \ .
\ea
Defining
\ba  \label{2G5}
\R_1(w_1,w_2) &:=& \sum_{p,q = 1}^{\infty}  k_{p+q}(H_1) w_1^{p-1}
w_2^{q-1}\nn\\
&=& \sum_{n=1}^\infty k_n(H_1) \sum_{p,q\ge 1\atop p+q = n} w_1^{p-1}
w_2^{q-1}\nn\\
&=& \sum_{n=1}^\infty k_n(H_1) \frac{w_1^n - w_2^n}{w_1 - w_2}
\ea
yields with (\ref{1G10}) the eqs. (\ref{2G12}) and  (\ref{2G6}). \hfill $\Box$
$\\$

Following Wegner \cite{Weg} (his eqs. (4.18)-(4.21)) one easily shows that
eq. (\ref{2G6}) is consistent with the sum rule (\ref{2G10}).  Using this sum
rule
together with (\ref{2G12}) yields the relation
\ba
\R_1[G(z_1), G(z_2)]
 \sum_{s}  \G (r,s,s,r;z_1,z_2)\  =\  -\, \frac{R_1[G(z_1)] - R_1[G(z_1)]}{z_1
- z_2}
 \ea
 which closely resembles a Ward identity. Since $R_1[G(z_1)]$ is the
self-energy
 of the 1PG the lhs may be interpreted as a vertex function.  The structure of
 eq. (\ref{2G6}) permits still another characterization
 of $\R_1[G(z_1), G(z_2)]$ as a local effective electron-electron interaction.
This  becomes particularly  pronounced if one represents
 (\ref{2G6}) diagrammatically by
 \begin{center} \fbox{here: Diagram 2} . \end{center}
 Thus eq. (\ref{2G6}) has a ladder structure with an effective
electron-electron
 interaction $\R_1[G(z_1), G(z_2)]$. This interaction  is a contact
interaction, i.e.,
  after averaging,  the two electrons
 propagate independently through the lattice unless they meet at
 same site.  As for 1PG our model
 resembles CPA character. We come back to this
 in sect. \ref{S4}.

\subsection{Long-range behaviour and conductivity} \label{S34}
\indent {\it Long-range behaviour.}
To discuss the long-range behaviour of the 2PG let us consider its connected
part for $r=r'$ and $s=s'$
\ba  \label{C1}
C(r,s;z_1,z_2) := \G (r,s,s,r;z_1,z_2) - G(r,s;z_1) G(s,r;z_2)
\ea
and its Fourier transform
\ba \label{C2}
\tilde C(q;z_1,z_2) = \sum_r C(0,r;z_1,z_2)\, \e^{iqr}\ .
\ea
The disconnected part of the 2PG yields only a short-range contribution
and  will not be discussed in the following.

Defining  $\G_{12}(r,s;z_1,z_2) := G(r,s;z_1) G(s,r;z_2)$ and its
 Fourier transform
\ba  \label{C3}
\tilde{\G}_{12} (q;z_1,z_2) &:=& \sum_{r} G(r,0;z_1) G(0,r;z_2)\, \e^{iqr}\nn\\
&=& \int_{q'} \tilde G(q';z_1) \tilde G(q'-q;z_2)
\ea
with $\tilde G(q;z)$ given by (\ref{1G20}), and using  (\ref{2G6}) one obtains
with the abbreviation $\R_1 \equiv  \R_1[G(z_1),G(z_2)]$
\ba
\tilde C(q;z_1,z_2) =  \R_1 \Bigm(\tilde{\G}_{12} (q;z_1,z_2)\Bigm)^2
 + \ \R_1\, \tilde{\G}_{12} (q;z_1,z_2) \tilde C(q;z_1,z_2)\ ,
 \ea
 which has the solution
\ba  \label{C4}
\tilde C(q;z_1,z_2) = \frac{ \R_1 \Bigm(\tilde{\G}_{12} (q;z_1,z_2)\Bigm)^2}{1
-
\R_1 \tilde{\G}_{12} (q;z_1,z_2)}\ .
\ea

Now we want to  show that $\tilde C(0;z_1,z_2)$ diverges  if $z_1$ and $z_2$
approach
the same energy $E$ from different halves of the complex plane along the
branch cut of $G$, that is for
\ba \label{taz}
\varrho(E) \equiv \frac{1}{2\pi i} \bigm[ G(E-i0^+) - G(E+i0^+)\bigm]\
\neq \ 0\ .
\ea
This follows from the decomposition
\ba
\tilde G(q;z_1) \tilde G(q;z_2) =
\frac{\tilde G(q;z_2) - \tilde G(q;z_1)}{z_1 - R_1[G(z_1)] -
z_2 + R_1[G(z_2)]}\ ,
\ea
where we have used (\ref{1G20}), which yields
\ba  \label{C5}
\tilde{\G}_{12} (0;z_1,z_2) = \frac{1}{\R_1} +
 \frac{z_2 - z_1}{\R_1 \bigm(z_1 - z_2\bigm) -
  \R_1^2\bigm(G(z_2) - G(z_1)\bigm)}\ .
\ea
Thus the denominator of (\ref{C4}) vanishes for $q = 0$ and lim$z_{1/2} = E$
since  the second term of
(\ref{C5}) vanishes because of (\ref{taz}).
{}From this one concludes that $\tilde C(q;z_1,z_2)$ has a diffusive pole.

To make this explicit we take
\ba
z_1 = E + \frac{1}{2} \omega\quad,\qquad  z_2 = E - \frac{1}{2} \omega
\ea
where $E$ is real and $\omega$ has an imaginary part of sign $s$.
Inserting this in (\ref{C5}) and assuming as Wegner \cite{Weg} cubic
symmetry with coordination number $n$ one finds by expanding around
$q=0$ in leading order for small $\omega$ and $q^2$
\ba  \label{C6}
n \tilde C(q;z_1,z_2)  =  \left(-\frac{i\omega s}{2\pi \varrho(E)} + A
q^2\right)^{-1}
\ea
with
\ba \label{C7}
A &=& - \left(\frac{\mu (E)}{\varrho(E)}\right)^2
\left.\frac{\partial \tilde{\G}_{12} (q;z_1,z_2)}{\partial
q^2}\right|_{q=0}\nn\\
&=& \frac{1}{2d}\left(\frac{\mu (E)}{\varrho(E)}\right)^2
\sum_r r^2 G(0,r;z_1) G(r,0;z_2)\ .
\ea
Here we have used (\ref{2G12}) to write
\ba
\R_1 [G(E+i0^+), G(E-i0^+)]  =
\frac{R_1[G(E-i0^+)] - R_1[G(E+i0^+)]}{G(E-i0^+) - G(E+i0^+)}
= \frac{\mu (E)}{\varrho (E)}
\ea
with the measure of $R_1[G(z)]$
\ba \label{C77}
\mu (E) := \frac{1}{2\pi i} \bigm[ R_1[G(E-i0^+)] - R_1[G(E+i0^+)]\bigm] \ .
\ea
In the case of the gaussian ensemble  with covariance $M$  we will show in
Sect. \ref{S5}
that $\mu (E) = M \varrho (E)$, i.e. $\R_1 \equiv M$, so that we find Wegner's
result
\cite{Weg} (his eqs. (4.40)-(4.41)). Thus eq. (\ref{C6})
merely differs from the
corresponding solution for the gaussian ensemble
 by a redefinition  of the constant $A$.
The function $\tilde C(q;z_1,z_2)$  essentially determines the long-range
and the $\omega\to 0$-limit of the 2PG. Without further  calculations one
concludes
from this that the qualitative  behaviour of the long-range- and the
$\omega\to 0$-limit of the 2PG does not depend on the distribution of the
disorder
in the electronic level space.

In more detail, following Wegner \cite{Weg} (his Sect. V), one sees from
(\ref{C6}) with the definition of the wave vector
\ba
\kappa = \left(\frac{-i\omega s}{2\pi A \varrho(E)}\right)^{1/2},\quad
         \mbox{Re}\,\kappa > 0
\ea
that eigenstates separated by an energy difference $\omega$ are correlated in
phase
over a length
\ba
L = |\kappa|^{-1} = [2\pi A \varrho(E)/|\omega|]^{1/2}
\ea
which diverges as $|\omega|^{-1/2}$ as $\omega\to 0$. Furthermore,
by Fourier back transformation and dimensional analysis one finds for fixed
$r\ll L$  and energies in opposite halves of the complex plane differing by
$\omega$
that $C(0,r;z_1,z_2)$ approaches a finite value provided $d>2$, diverges
logarithmically  as a function of $\omega$ for $d=2$ and like
$|\omega|^{d/2 - 1}$
for $0\le d<2$, respectively, for $\omega\to 0$. This implies that for $d>0$
the eigenstates are extended since $C$ does not diverge as fast as
$|\omega|^{-1}$. Up to  a redefinition of the constant $A$ this is the
same result as for the gaussian ensemble \cite{Weg}.
$\\$

{\it Conductivity.}
The 2PG determines the conductivity $\sigma_{_T} (\omega)$
via the Kubo-Greenwood  relation \cite{KP}
\ba  \label{C01}
\sigma_{_T}(\omega) = \frac{2 e^2}{\pi{\cal V}}
 \int_{-\infty}^{\infty} \omega^{-1}
\left[n_F\left(E-\frac{\omega}{2}\right) -
n_F\left(E+\frac{\omega}{2}\right)\right]
 \sigma(\omega,E) \d E \ .
\ea
Here,   $n_F(E) = [\exp\{ (E-E_F)/T\} + 1]^{-1}$
is the Fermi distribution, $E_F$ is the Fermi-energy, and
 $\sigma(\omega,E)$ is the current-current or the
density-density spectral function (cf. \cite{Weg}).

  At $T=0$  and in the dc limit $\omega  \to 0$ the conductivity is given by
  the spectral function itself
\ba  \label{C02}
\sigma_{_{T=0}}(\omega) = \frac{2 e^2}{\pi{\cal V}} \sigma(\omega,E_F)\ ,\qquad
\omega \to 0,
\ea
 where the spectral function $\sigma(\omega,E_F)$ is
 given by the connected part of the 2PG \cite{Weg} yielding
 \ba
\sigma_{_{T=0}}(\omega\to 0) &=& \frac{e^2 n}{4\pi^2{\cal V}} \omega^2
\sum_s \left.\frac{\partial}{\partial q^2}
\tilde C\left(q; E_F+\frac{\omega}{2} + is0^+,
 E_F-\frac{\omega}{2} + is0^+\right)\right|_{q=0}\nn\\
&=& \frac{2\pi e^2 n A}{{\cal V}}\, \varrho^2(E_F)\ .
\ea
Using the definition of $A$ in (\ref{C7}) and the Fourier transform
(\ref{1G20})
this result can also be quoted as follows
\ba
\sigma_{_{T=0}}(\omega\to 0) = \frac{2\pi e^2 n B}{{\cal V}}\, \mu^2(E_F)\ ,
\ea
where   $\mu (E_F)$ is
  the spectral function of $R_1[G(z)]$ (cf. (\ref{C77})) and
\ba
B := A \,\left(\frac{\varrho(E_F)}{\mu (E_F)}\right)^2 =  -\left.\frac{\partial
\tilde{\G}_{12} (q;z_1,z_2)}{\partial q^2}\right|_{q=0} \ .
\ea
The last two equations show that the dc conductivity
 at zero temperature is essentially given by the square of the spectral
function
of the Fourier transform of the 1PG. Again this result differs from that
for the gaussian ensemble by a mere redefinition of the constant $A$.
{}From this one concludes that the conductivity is nonvanishing
 everywhere inside the band, and that localization cannot occur in our model.

\setcounter{equation}{0}
\section{Coherent-Potential Approximation}  \label{S4}
One of the most effective approximation methods for the
Anderson model is the (single-site) {\it coherent-potential approximation}
(CPA)
initially proposed by Soven \cite{Sov} and Taylor \cite{Tay}.
Its main idea can be summarized as follows \cite{Econ,Lif}:
 One introduces an effective homogeneous
medium with the propagator  $G_0(r,s;z-\Sigma)$ with an effective potential
 $\Sigma$ in which the electron moves and demands
\ba  \label{CPA1}
G(r,s;z) = G_0(r,s;z-\Sigma(z))\ .
\ea
In other words, CPA calculates $G$
from an effective hamiltonian $H_{eff} = H_0
+ \sum_{r} \Sigma |r\rangle\langle r|$.  The coherent potential $\Sigma$ is
determined
in such a way that the difference  between the actual and the effective
hamiltonian,
$H-H_{eff}$,  produces on the average zero scattering at one site,
i.e. the averaged single-site $t$ matrix vanishes
\ba  \label{CPA2}
\langle t(z)\rangle :=
 \left\langle
\frac{f_r - \Sigma}{1 - (f_r -\Sigma) G(z)}\right\rangle = 0\ ,
\ea
where $G(z) \equiv G(r,r;z)$ is given by
\ba  \label{CPA1a}
G(z) = G_0(z-\Sigma(z))\ .
\ea
 Due to the translation invariance,
$t(z)$ is site-independent.
Velick\a'y \cite{Vel}
has worked out this concept for the 2PG and has found the following
CPA equation for the 2PG
\ba  \label{CPA2a}
\G (r,s,s',r';z_1,z_2) &=& G(r,s;z_1) G(s',r';z_2)\nn\\
&+&  \L(z,z')
 \sum_{r''} G(r,r'';z_1)  \G (r'',s,s',r'';z_1,z_2)   G(r'',r';z_2) \ ,
\ea
where $G$ is given by (\ref{CPA1}) with $\Sigma$ satisfying (\ref{CPA2}) and
\ba  \label{CPA2b}
\L(z,z') = \frac{\langle t(z) t(z')\rangle}{1 + G(z)\langle t(z) t(z')\rangle
G(z')}.
\ea
Here $\langle t(z) t(z')\rangle$ contains all contributions
from repeated scattering at the same site. Thus the 2PG is given as the sum of
single-site contributions in agreement with the general CPA philosophy.

In view of  (\ref{CPA1}) and (\ref{CPA2a})
the solutions of our model (\ref{1G12}) and (\ref{2G6}) have CPA character.
This has already been realized by Wegner \cite{Weg} and Khorunzhy and Pastur
\cite{KP} for the $n$-orbital
 model. Wegner showed that the $n\to\infty$ limit
of his model yields the CPA solution of the Anderson model provided that the
$f_r$ are
distributed according to the semi-circle law \cite{Weg}. In general the
connection between
CPA and models like Wegner's $n$-orbital model
is not clear (see the discussion
in \cite{KP}).

Here, we will show that the concept of freeness allows to put CPA on a firm
mathematical basis, namely, we can prove the following theorem:$\\$

{\sc Theorem} 4: Let the hamiltonian $H$ be given  by (\ref{I1})-(\ref{I3})
where
 $f_1,f_2,\ldots$ are {\it free} and identically  distributed according to
 a distribution $P$.
Then,   the solution for the
1PG given by Theorem 2 and the solution for the
2PG given by Theorem 3
 is identical to the CPA solution of
the  Anderson model where the $f_1,f_2,\ldots$ are {\it independent}
and identically distributed with the same  $P$. \\
\indent
{\sc Proof}: We first prove the assertion of the theorem
for the 1PG. We show that (\ref{1G12}) and
(\ref{1G9}) solve the CPA equations (\ref{CPA1},\ref{CPA2}) if we identify
  $\Sigma(z) = R_1[G(z)]$. Using this to  rewrite the CPA condition
(\ref{CPA2}) as
\ba  \label{CPA5}
\langle t(z)\rangle  =  \sum_{n=1}^\infty \langle (f_r - R_1(w))^n\rangle
w^{n-1}
\ea
where  $w:= G(z)$,
one finds
\ba
\langle t(z)\rangle  + \frac{1}{w}
 &=& \sum_{n=0}^\infty \langle (f_r - R_1(w))^n\rangle  w^{n-1}\nn\\
&=& \sum_{n=0}^\infty \sum_{k=0}^n {n\choose k}  \langle f_r^k\rangle
       \Big( - R_1(w)\Big)^{n-k}  w^{n-1}\nn\\
&=& \sum_{k=0}^\infty \langle f_r^k\rangle
        \left( \sum_{n=k}^\infty {n\choose k}  \Big( - w R_1(w)\Big)^{n-k}
\right)  w^{k-1}\nn\\
&=&  \frac{1}{w^2}
        \sum_{k=0}^\infty \frac{\langle f_r^k\rangle}{\Big(R_1(w) +
1/w\Big)^{k+1}} \nn\\
&=& \frac{1}{w^2} \  G_1[R_1(w) + 1/w]\nn\\
 &=& \frac{1}{w}\ ,
\ea
i.e. $\langle t(z)\rangle = 0$. Here, we have used the
equivalent form of (\ref{1G9}), $G_1[R_1(w) + 1/w] = w$,
 and the identity
\ba
\sum_{n=k}^\infty {n \choose k} a^{n-k} = \frac{1}{(1-a)^{k+1}}\ .
\ea
To prove the assertion of the theorem
for the 2PG we show that $\L(z,z') = \R_1[G(z),G(z')]$. We first calculate the
average
of the product of the single-site t matrices
\ba
 \Lambda (z,z') := \langle t(z) t(z') \rangle = \left\langle
\frac{f_r - \Sigma}{1 - (f_r-\Sigma) G}
\frac{f_r - \Sigma'}{1 - (f_r-\Sigma') G'}\right\rangle\ ,
\ea
where we have used the abbreviations
$G = G(z)$, $G' = G(z')$, $\Sigma = \Sigma(z)$, $\Sigma' = \Sigma(z')$.
The average can be evaluated using (\ref{CPA2}).  Identifying
$\Sigma(z) = R_1[G(z)]$ one finds
\ba
 \Lambda (z,z') = \frac{R_1[G] - R_1[G']}{G - G' + G G'(R_1[G'] - R_1[G])}
\ea
which reduces with eq. (\ref{2G12}) to
\ba \label{CPAx1}
\Lambda (z,z') = \R_1[G,G'] / \big(1-\R_1[G,G'] G G'\big)\ .
\ea
Solving this equation for $\R_1$ yields the rhs of (\ref{CPA2b})
and therfore proves the second assertion of our theorem. \hfill $\Box$ $\\$

In CPA both the 1PG and the 2PG are entirely determined by the coherent
potential
$\Sigma(z)$ which follows  self-consistently from (\ref{CPA2}) and
(\ref{CPA1a}). In our model $\Sigma(z)$  is given by Voiculescu's
 $R$-transform $R_1$ of
the disorder. Thus the prescription (\ref{1G9}) for calculating $R_1$
can be considered as the formal solution of the CPA equations for arbitrary
disorder.

\setcounter{equation}{0}
\section{Discussion of specific distributions} \label{S5}
 \subsubsection{Deterministic noise}
  Let us start with the trivial case
where all $f_r$ are deterministic attaining the constant $\gamma$, i.e. they
have a $\delta(f_r - \gamma)$-distribution for all $r$. Then, $G_1(z) =
[z - \gamma]^{-1}$, hence $R_1(w) = \gamma$ (cf. eq. (\ref{1G9})),  hence
$\R_1(w_1,w_2) = 0$ (cf. eq. (\ref{2G12})).  This yields the following solution
for the
Green functions
\ba  \label{D1}
G(z) &=& G_0(z-\gamma)\nn\\
\tilde G(q;z)&=& \frac{1}{z-\tilde{v}(q)-\gamma}\\
\G (r,s,s',r';z_1,z_2) &=& G(r,s;z_1) G(s',r';z_2)\nn ,
\ea
i.e. the connected part  $C$ of the 2PG vanishes.
Since for deterministic  $f_r$ there is no difference between independence
and freeness, this is also the exact solution of the original Anderson model.
Of course it is just given by a trivial energy shift $\gamma$ of the
unperturbed solution.

 \subsubsection{Cauchy (Lorentz) noise: The Lloyd model}
Consider now the Lloyd model
 where the $f_r$ are distributed according to a Cauchy-distribution with
parameter
 $\gamma$, i.e.
 \ba  \label{D2}
 \d P(f_r = \epsilon) = \frac{1}{\pi} \frac{\gamma}{\gamma^2 + \epsilon^2}
 \d \epsilon\ .
 \ea
 Note that moments, and thus also cumulants
of $H_1$ do not exist in this case,
 but nevertheless our main formulas for the connection between
 $G_1(z)$ and $R_1(w)$  can be
 justified in this case too \cite{Maa,BV}.
 We have
 \ba  \label{D3}
 G_1(z) = \frac{1}{\pi}\int_{-\infty}^{\infty} \frac{\gamma}{\gamma^2 +
 \epsilon^2}
 \frac{1}{z-\epsilon} \d \epsilon = \frac{1}{z+ i\gamma}\ ,
 \ea
 hence $R_1(w) = is\gamma$ with $s$ being the sign of the imaginary part of
$w$.
 Using (\ref{2G12}) one sees that
 $\R_1(w_1,w_2) = 0$ if $w_1$ and $w_2$ are on the same halves of the complex
plane, and
 \ba
  \R_1(w_1,w_2) = \frac{2is_1\gamma}{w_1 - w_2}
  \ea
   if $w_1$ and $w_2$ are on opposite halves.  Thus in the former case,
   we find the same result as in (\ref{D1})
 with $\gamma$ replaced by $i\gamma$.
 In the latter case the connected part $C$ of the 2PG does not vanish so
 that one finds  a finite conductivity
 \ba
\sigma_{_{T=0}}(\omega\to 0) = \frac{2 e^2 n B}{\pi {\cal V}}\, \gamma^2\
\ea
with $B = - (\partial \tilde{\G}_{12} (q)/\partial q^2)|_{q=0}$.

 \subsubsection{Gaussian random matrix  noise: The Wegner model}
  Wegner's model
 consists in choosing the $f_r$ to be  --
in the limit $n\to\infty$ --  symmetric
 $n\times  n$-gaussian random matrices with the entries
 of $f_r$ and of $f_{r'}$ being
 independent for $r\neq r'$.  As we have explained in Sect. \ref{S21} and
\ref{S31}   this means nothing
 but that the $f_r$ are free. Thus Wegner's model is the special case of ours
 where the distribution of the $f_r$ is given by the
 eigenvalue disribution of symmetric gaussian random matrices, i.e. by Wigner's
 semi-cirle law,
 \ba  \label{D5}
  \d P(f_r = \epsilon) =  \frac{1}{2\pi M} \sqrt{4M - \epsilon^2} \d \epsilon
  \ea
for $\epsilon^2 \le 4 M$ and zero elsewhere.
 The fact that gaussian random matrices are free explains
quite naturally Wegner's observation that his model gives the same result
as the CPA with a semi-circle distribution applied to the Anderson model.

For the semi-circle law one has
\ba  \label{D6}
G_1(z) = \frac{z-\sqrt{z^2 - 4 M}}{2 M} = \frac{1}{z- M G_1(z)}\ ,
\ea
which yields $R_1(w) = M w$. This means that only the second non-crossing
cumulant is different from zero and thus we have $\R_1(w_1,w_2) =  M$.
This gives the following solution
\ba  \label{D7}
 G(z) &=& G_0\Big[z- M G(z)\Big]\\
       \label{D8}
\tilde G(q;z)&=& \frac{1}{z-\tilde v(q)- M G(z)}\\
       \label{D9}
\tilde{C} (q;z_1,z_2) &=&
\frac{ M \Bigm(\tilde{\G}_{12}(q;z_1,z_2)\Bigm)^2}{1 -  M  \tilde {\G}_{12}
(q;z_1,z_2)}\\
       \label{D10}
\sigma_{_{T=0}}(\omega\to 0) &=& \frac{2\pi e^2 n A}{{\cal V}}\, \varrho^2(E)
\ea
with $A = -M^2 (\partial \tilde{\G}_{12} (q)/\partial q^2)|_{q=0}$.
These formulas were  found by Wegner \cite{Weg} and later rederived by
 Khorunzhy and Pastur \cite{KP}.
  The deformed semi-circle law
(\ref{D7}) had also appeared earlier in the work of Pastur \cite{Past1} as the
solution for the problem of determing the eigenvalue distribution of a sum
$W + D$ of a symmetric gaussian random matrix $W$ and a non-random diagonal
matrix $D$. By our remarks in Sect. \ref{S2},
this latter problem is nothing but
calculating the free  convolution of the distribution of $W$ and of $D$ and
hence, in the light of Theorem 1, the
coincidence of Wegner's and Pastur's result appears
as no surprise.

One should also note that in the context of the
free convolution the semi-circle distribution plays  the same role as the
gaussian distribution for the classical
convolution. This can be seen, for instance, from the fact that only the second
non-crossing cumulant is different from zero for the  semi-circle distribution,
similiarly as only the second usual cumulant is non-vanishing for the
gaussian distribution. For more details on the  ``free gaussian" and related
topics, like free central limit theorem or free Poisson law, we refer to
\cite{VDN,Spe3,Maa}. $\\$

\subsubsection{q-noise: An interpolation}
In \cite{NSp2}
we have introduced a new class
of stochastic processes which interpolate continuously
between classical,  gaussian,  random matrix,
dichotomic, and Poisson processes.
  This construction can be adapted for quenched multi-site
disorder as follows: Let the disorder at each site $r$ be given by
\ba \label{D11}
 \hat{f}_r \;:=\; \sigma\, (\,a_r \; +\;  a_r^{\dag}\,)\
 + \   \xi\, a_r^{\dag} a_r
\ea
 in terms of  deformed annihilation and creation operators $a_r$,
  $a_r^{\dag}$ on some Hilbert space $\H$.
 These operators satisfy at each site $r$
the deformed canonical commutation relations
\ba \label{D12}
a_r \,a_{r}^{\dag} \  -\    q \,a^{\dag}_{r}\,a_r &=&  {\bf 1}  \ ,\\
a_r|0\rangle &=& 0\ ,
\ea
where ${\bf 1}$ and $|0\rangle$ denote the identity operator and the vacuum in
$\H$,
respectively.
At different sites $r\neq r'$ the operators are assumed to be free, implying
 that $\hat f_1,\hat f_2,\ldots$ are free, i.e.
\ba \label{D13}
a_r \,a_{r'}^{\dag}  \equiv 0   \ , \qquad r \neq  r'\ .
\ea
  The deformation   parameter $q$ is
  real and  varies continuously in the interval $-1\ \le \ q\ \le\ 1$.
  For $\xi = 0$, the limiting cases $q=1$ and $q=-1$  describe  gaussian and
  dichotomic disorder, respectively,
  whereas the case $q=0$  corresponds to
  Wegner's $n$-orbital model.  The $\xi\, a_r^{\dag} a_r$-term allows to
include
  Poisson-like disorder \cite{NSp2,HatPat}

Eq. (\ref{D13}) is an alternative representation of those free random variables
which can be represented   by deformed creation and annihilation operators.
To make the construction clear, let us formulate the original Anderson model
for gaussian site-diagonal disorder ($q=1$, $\xi = 0$)
in  this language: $\hat{f}_r := \sigma (a_r + a_r^{\dag})$
 with $a_r a_{r'}^{\dag}  -  a^{\dag}_{r'} a_r  =  \delta_{r,r'}{\bf 1}$ for
all $r,r'$, in particular, $a_r a_{r'}^{\dag} =  a_{r'}^{\dag} a_r$ for
$r\neq r'$,  which is
 clearly different from (\ref{D13}).

  We can now identify moments of the random variables $f_r$
in (\ref{I3}) with the
  Hilbert space vacuum  expectation values of
products of $\hat{f}_r$ by means of a
  generalized Wick Theorem \cite{NSp2,FB}.
By using the partial cumulants we have
  calculated the 1PG of $H_1$ (cf. eq. (36) in \cite{NSp2})
\ba  \label{D14}
G_1(z) &= & \frac{1}{z-R_1[G_1(z)]}\nn\\
&=&  \frac{1}{\displaystyle z \,- \,
\frac{\sigma^2 q^{(0)}}{\displaystyle z \,-\,   \xi q^{(0)} \,- \,
\frac{\sigma^2 q^{(1)}}{\displaystyle z \,-\,   \xi q^{(1)} \,- \,
\frac{\sigma^2 q^{(2)}}{\displaystyle z \,-\,   \xi q^{(2)}\,- \,
\frac{\sigma^2 q^{(3)}}{\displaystyle \ddots}}}}} \nn\\
&=:&  1/\Big(z \,- \, \Big(
\sigma^2 q^{(0)} /\Big(z \,-\,   \xi q^{(0)} \,- \,
 \Big(
\sigma^2 q^{(1)} / \Big(z \,-\,   \xi q^{(1)} \nn\\
&&  \qquad  - \,
\Big(
\sigma^2 q^{(2)} / \Big(z \,-\,   \xi q^{(2)}\, - \,
\Big(
\sigma^2 q^{(3)} /  \ldots \Big)\ldots \Big)
\ea
where
\ba  \label{D15}
q^{(k)}:= 1 + q + q^2 + \ldots + q^k = \frac{1-q^{k+1}}{1-q}\ .
\ea
Thus,
\ba \label{D16}
R_1[G(z)] &=& G(z) \sigma^2 q^{(0)} /
 \Big(1 + G(z)( R_1[G(z)] -   \xi q^{(0)})  \nn\\
&+& \Big(
G(z)^2 \sigma^2 q^{(1)} /
\Big(1 + G(z)(R_1[G(z)]  -   \xi q^{(1)})  \nn\\
&+& \Big(
G(z)^3 \sigma^2 q^{(2)} /
\Big(1 + G(z)(R_1[G(z)]  -   \xi q^{(2)})   \nn\\
&+& \Big(
G(z)^4 \sigma^2 q^{(3)} /  \ldots\Big) \ldots \Big)
\ea
which  together with
\ba
G(z) = G_0\Big[z-R_1[G(z)]\Big]
\ea
is a closed set of  nonlinear self-consistent equations for the
1PG and $R_1(w)$.
The continued fraction
(\ref{D16}) can be summed in closed form
for $q=-1$, where  one finds, for $\xi = 0$,
$R_1[G(z)] = (\sqrt{1 + 4\sigma^2G^2(z)} - 1)/2G(z)$, and
for $q=0$, where  one finds, for $\xi = 0$,
$R_1[G(z)] = \sigma^2 G(z)$ and thereby Wegner's model with $\sigma^2 = M$.

\setcounter{equation}{0}
\section{Summary} \label{S6}
In this paper we have applied the concept of free random variables,
invented by Voiculescu  in a mathematical context, to the
 tight-binding hamiltonian
with site-diagonal disorder of an electron in a periodic solid.
The difference of our model to the usual Anderson model
lies in the fact  that instead
of assuming the disorder to be independent at different lattice sites, we
have assumed it to be free.

 Both freeness and independence  can be considered
as a rule  for calculating mixed moments of random variables. In contrast to
the  case of independent  disorder,  free noise does imply a treatable
relation between $H_0$ and $H_1$: they are also free. This finally
allows to close the infinite hierachy  of equation of motion and
to calculate  all physically relevant quantities.

In sect. \ref{S3} we have argued that in the limit $n\to\infty$
free noise can be represented
by $n\times n$-random matrices which are randomly rotated
against each other at different lattice-sites, starting from a matrix with
fixed
but arbitrary eigenvalue distribution in the limit $n\to\infty$.
 In this sense,
our model is an extension of Wegner's $n$-orbital model for the gaussian
ensemble
to arbitrary eigenvalue distribution in the energy level space.
 One of the most striking property of our model  is
  that both the long-range behaviour
and the zero-frequency limit of the 2PG are universal with respect
to the eigenvalue distribution in the energy level space.

One surprising feature of the Wegner model is that its solution
coincides with a special CPA solution. This generalizes also to
our model. In sect. \ref{S4} we have shown
that our solution for the 1PG and the 2PG
 also solves the CPA equations for the Anderson model with the same
distribution
 of disorder.
Note that we specify our
model rigorously in the
beginning and that we are able to calculate all quantities without
any further approximation. Thus our multi-site model is a
rigorous mean field model
for the usual single-site CPA.
The $R$-transform of Voiculescu, $R_1$,  may be considered as
the formal solution of the CPA equations for arbitrary disorder.
It posseses the physical interpretation as an effective local potential;
the corresponding quantity for the 2PG, $\R_1$, can be considered
as an effective local electron-electron interaction.
Cleary as seen in sect. \ref{S5},
aside from some specific distributions, both functions cannot be calculated
analytically. However, due to the mentioned universality,  the Wegner
model is exemplary within many respects and it might be sufficient to restrict
oneself to this case  in the general frame of a mean field approximation.

 Furthermore, our description using the theory of free
random variables and the notion of non-crossing cumulants
allows a straightforward generalization to the case of dynamical
disorder and thus promises to give a
rigorous model for dynamical CPA. These
subjects will be
pursued further in forthcoming investigations.

\acknowledgements
Helpful discussions with Dr. Petr Chvosta and Dr. Reimer K\"uhn
 and valuable comments of
 Prof. Dan Voiculescu are gratefully acknowledged. We would  particularly
like to thank  Prof. Franz Wegner for his insightful suggestions and his
carefully reading of the manuscript.
This work was partly supported by the Deutsche Forschungsgemeinschaft (R.S.).

\end{document}